\documentclass[twocolumn]{aastex62}

\newcommand{\msun}{\mathrm{M}_\odot}
\newcommand{\ce}{\mathrm{CE}}
\newcommand{\nj}{\mathrm{Nanjing}}
\newcommand{\pgal}{\mathrm{galaxy}^{-1}}
\newcommand{\pmyr}{\mathrm{Myr}^{-1}}
\newcommand{\pgpcc}{\mathrm{Gpc}^{-3}}
\newcommand{\pyr}{\mathrm{yr}^{-1}}
\newcommand{\psec}{\mathrm{s}^{-1}}

\newcommand{\sdeg}{\mathrm{deg}^2}
\usepackage{amsmath}

\received{xxx}
\revised{yyy}
\accepted{zzz}
\submitjournal{ApJ}

\shorttitle{rate of iptf 14gqr like ultra-stripped sne}
\shortauthors{Hijikawa et al.}

\begin{document}

\title{The rate of iPTF 14gqr like ultra-stripped supernovae and binary evolution leading to double neutron star formation}

\correspondingauthor{Kotaro Hijikawa}
\email{hijikawa@astron.s.u-tokyo.ac.jp}

\author{Kotaro Hijikawa}
\affil{Department of Astronomy, Graduate School of Science, the University of Tokyo, Tokyo 113-0033, Japan}

\author{Tomoya Kinugawa}
\affiliation{Department of Astronomy, Graduate School of Science, the University of Tokyo, Tokyo 113-0033, Japan}

\author{Takashi Yoshida}
\affiliation{Department of Astronomy, Graduate School of Science, the University of Tokyo, Tokyo 113-0033, Japan}

\author{Hideyuki Umeda}
\affiliation{Department of Astronomy, Graduate School of Science, the University of Tokyo, Tokyo 113-0033, Japan}



\begin{abstract}
Double neutron star (DNS) systems are produced from massive binaries. A supernova (SN) explosion of an extremely stripped star is expected to occur at the final stage of DNS formation. This type of SN is called an ultra-stripped SN (USSN). Recent research revealed that a type Ic SN, iPTF 14gqr (SN 2014ft), has low ejecta mass ($\approx 0.2~\mathrm{M}_\odot$) and its progenitor has a helium envelope with its mass $\sim 0.01~\mathrm{M}_\odot$. This SN is interpreted as a USSN, and thus this is the first discovery of a USSN. Furthermore, the observation of iPTF 14gqr provides us with some information about its formation history. Here, we perform rapid population synthesis calculations so as to estimate the detection rate of iPTF 14gqr like USSNe with optical transient surveys: the intermediate Palomar Transient Factory (iPTF), the Zwicky Transient Facility (ZTF), and the Large Synoptic Survey Telescope (LSST). We find that iPTF, ZTF, and LSST can observe iPTF 14gqr like USSNe at the rates of 0.3, 10, and 1 $\mathrm{yr}^{-1}$, respectively. The iPTF can detect 1 iPTF 14gqr like USSN during its four year observation. We also investigate effects of mass-loss efficiency during Roche-lobe overflow on formation channels.
\end{abstract}

\keywords{binaries: general - stars: evolution - supernovae: general}


\section{Introduction\label{sec:intro}}
The first gravitational-wave (GW) signal from a double neutron star (DNS) merger, GW170817 was recently detected \citep{abbott17c}. DNS systems play important roles in various fields of astronomy and astrophysics. For example, using GW signals, we can constrain the equation of state of a highly dense nuclear matter \citep{annala18}. DNS merger events are suggested to lead to short gamma-ray bursts \citep{eichler89}. The merger events are also expected to be a place of {\it r}-process nucleosynthesis \citep[e.g.,][]{rosswog15}.

The canonical formation channel from binary zero-age main-sequence (ZAMS) systems with their masses $\sim 10~\msun$ to DNS systems have been proposed by many authors \citep[e.g.,][]{bhattacharya91,tauris06}. The overview of the canonical channel is as follows \citep[see Figure 1 in ][]{tauris17}. At first, a mass transfer called Roche-lobe overflow (RLO) from the primary to the secondary occurs, then the primary causes supernova (SN) explosion and a neutron star (NS) is born. Next, this system becomes a high-mass X-ray binary (HMXB) and the NS plunges into the envelope of the secondary, then the hydrogen envelope is ejected. This is called the common envelope (CE). Then this system turns into an NS--helium star binary, and case BB RLO from the secondary to the NS occurs. Finally, the secondary causes SN explosion and a second NS is born. After that, the orbit of this system is circularized and shrunk due to the GW radiation. 

A system has to experience SN explosion twice until the formation of the DNS system, and at that time, a pulsar kick is usually imparted on a newborn NS. Pulsar kicks come about as a result of nonradial hydrodynamic instabilities. If a pulsar kick velocity is large enough, this system disrupts. To form DNS systems, especially systems that will merge within a Hubble time, an SN explosion with smaller kick is essential. \citet{tauris15} propound that an extremely stripped progenitor can explode with smaller kick for the following two reasons. First, the binding energy of the stripped envelope is very small ($\sim 10^{49}~\mathrm{erg}$), so that a relatively weak shock can eject this envelope before an anisotropy grows. Second, since the ejecta of the extremely stripped star is lighter ($\sim 0.1~\msun$) than a canonical SN ($\sim1$ to 10 $\msun$), the gravitational tug from the ejecta to the proto NS (PNS) is weaker. Actually, \citet{suwa15} and \citet{muller18} perform multidimensional SN simulation and report that these extremely stripped progenitors explode with small kicks $\sim\mathcal{O}(10)~\mathrm{km~s^{-1}}$. This type of SN whose helium envelope is extremely stripped by binary interactions is called an ultra-stripped SN (USSN) \citep{tauris13,tauris15}. Note that \citet{tauris17} suggest that some USSNe producing massive NSs receive larger kicks.

Therefore, a USSN is likely to occur in the final stage of DNS formation. Recently, it was found that a type Ic SN iPTF 14gqr (SN 2014ft) has an ejecta mass $M_\mathrm{ej}\approx 0.2~\msun$ and helium envelope mass $M_\mathrm{He,env}\sim 0.008~\msun$; thus, it can be interpreted as a USSN \citep{de18}. These facts were obtained thanks to the observation of rapid decline of the first peak due to the shock cooling emission and second peak by the radioactive decay. Furthermore, the observation of iPTF 14gqr gave us suggestions about its formation history, i.e., the binary evolution. This is the first ever detection of a USSN. Type Ic SNe such as SN 2005ek \citep{drout13} and SN 2010X \citep{kasliwal10} show fast evolving and low peak ($\lesssim 10^{42}~\mathrm{erg~s^{-1}}$) light curves, so that these are suggested to be classified as USSNe. However, since these events were observed only at the radioactively powered peak, the origins of these SNe remain uncertain. Thus, iPTF 14gqr is considered to be the first discovered USSN. The observability of helium of USSNe is also argued by \citet{moriya17}.

In this study, we estimate the rate of iPTF 14gqr like USSNe by the rapid population synthesis method. Then we discuss the observability of iPTF 14gqr like USSNe for optical transient surveys the intermediate Palomar Transient Factory (iPTF), the Zwicky Transient Facility (ZTF), and the Large Synoptic Survey Telescope (LSST). Additionally, we roughly assume that merging DNS systems involve one USSN and estimate the detection rate from other authors' DNS merger rates. We also argue the massive binary evolution and formation channels leading to DNS systems. 

Various population synthesis studies were previously conducted to estimate the merger rate of binary systems that are comprised of an NS or a black hole (BH) in the local universe \citep[e.g., ][]{andrews15,belczynski18,chruslinska18,giacobbo18,kruckow18,shao18,vigna18}. On the other hand, we focus on USSNe and their detection rates. Later in this paper (Section 4.1.), we compare their results with ours. This paper is organized as follows. In section \ref{sec:method}, we describe population synthesis method and adopted input physics. The results of our calculation and estimated detection rate of iPTF 14gqr like USSNe are in Section \ref{sec:results}. We discuss effects of some binary physics on formation channels, and the merger rate of DNS systems in Section \ref{sec:discussion}. Finally, we summarize this study in Section \ref{sec:summary}.


\section{Method\label{sec:method}}
To estimate the rate of USSNe, we perform rapid binary population synthesis calculation. We use a binary population synthesis code based on the BSE code \citep{hurley02}. We rewrite a mass transfer timescale, a merger criterion, the binding energy parameter $\lambda$, and responses of the radius of the donor star to an adiabatic mass change. We will explain these in detail later. In this study, we calculate $10^6$ binaries consisting of ZAMS stars with the solar metallicity $Z_\odot=0.02$. We randomly determine the initial primary mass with the Salpeter initial mass function \citep{salpeter55}. Then we use the flat distribution for the mass ratio $q\equiv M_2/M_1$ distribution ($M_1$, $M_2$ are the mass of the primary and the secondary star, respectively) to determine the initial secondary mass \citep{kobulnicky07,kobulnicky12}. We assume that the initial separation follows the log-flat distribution \citep{abt83}, and the initial eccentricity is zero. We adopt a constant star formation rate $2.0~\msun~\pyr$ following the Galactic value $1.9\pm0.4~\msun~\pyr$ \citep{chomiuk11}.

A massive binary evolution has large uncertainties, for example, in a mass transfer phase. Therefore, we use various models whose input physics are different from each other. We consider four values for the mass-loss efficiency $\beta$, and the two cases: if a binary system coalesces when the system enters the CE phase while the donor star is in the Hertzsprung gap (HG) or if it does not. Although we use $\lambda_\nj$ \citep{xu10,xu10e} as a binding energy parameter for CE in the above models, we additionally consider three fixed values for $\lambda$ to compare our results with previous population synthesis studies. 
More detailed descriptions of the input physics are presented in Section \ref{subsec:rlo}--\ref{subsec:sn}. For each model, we calculate a probability distribution of the orbital parameters of DNS systems. Then we evaluate the likelihood of each probability distribution using the observed 15 Galactic DNS systems as samples. This calculation method follows \citet{andrews15} and \citet{vigna18}. With the maximum likelihood model (i.e., the most favored model), we estimate the rate of iPTF 14gqr like USSNe (which will be defined in Section \ref{subsec:sn}). Additionally, we evaluate the formation rate and merger rate of DNS systems, and discuss preferable input physics to explain the orbital parameters of the observed DNS systems.


\subsection{Roche-lobe overflow\label{subsec:rlo}}
When a star in a binary system expands and fills its Roche lobe, a mass transfer occurs. If the mass transfer is dynamically unstable, the envelope of the donor star engulfs the companion star. Then the companion star expels the envelope of the donor star, and finally the binary system merges or becomes a close binary. This process is called CE. On the other hand, the mass transfer that is stable on a dynamical time-scale but unstable on a thermal time-scale is called RLO. We assume that RLO always occurs on a thermal time-scale (approximately equal to the Kelvin-Helmholtz time-scale $\tau_\mathrm{KH}=GMM_\mathrm{env}/RL$, where $G$ is the gravitational constant, $M$ is the mass of the donor star, $M_\mathrm{env}$ is the envelope mass of the donor star, $R$ is the radius of the donor star, and $L$ is the luminosity of the donor star). In particular, CE and Case BB RLO, which a helium star after CE phase is expected to experience, are important processes of the stripping of the exploding star \citep{tauris13,tauris15}. We use the helium star evolution given in \citet{hurley00}.

In our calculation, to discriminate CE from RLO, we compare $\zeta_\mathrm{ad}$ with $\zeta_\mathrm{L}$. $\zeta_\mathrm{ad}\equiv(\mathrm{d}\ln R_\mathrm{L}/\mathrm{d}\ln M)_\mathrm{ad}$, which is the response of the radius of the donor star to an adiabatic mass change, and $\zeta_\mathrm{L}\equiv\mathrm{d}\ln R_\mathrm{L}/\mathrm{d}\ln M$, which is the response of the Roche-lobe radius to a mass change. If $\zeta_\mathrm{L}>\zeta_\mathrm{ad}$, CE occurs, and otherwise RLO occurs. The value of $\zeta_\mathrm{ad}$ depends on the structure of a star, or the evolutionary phase. We use fixed values $\zeta_\mathrm{ad}=1.95,~5.79$ for helium main-sequence stars, and helium giant stars, respectively \citep{ivanova02,belczynski08}. For the other evolutionary types, we follow the BSE code.

The parameter $\beta$ that represents how many masses are lost from a binary system by RLO (i.e., $\mathrm{d}M_1=-(1-\beta)\mathrm{d}M_2$, where $\mathrm{d}M_2(<0)$, $\mathrm{d}M_1(>0)$ is the net mass lost from the donor star, and accreted onto the companion star, respectively) has uncertainties. Furthermore, the value of $\beta$ is poorly constrained by observation \citep{demink07}. The larger the value of $\beta$, the more masses are lost from a binary system, so that the separation can become larger. Therefore, the parameter $\beta$ is important to determine the binary ultimate fate. In this study, we assume that $\beta$ is constant through the evolution and consider four different values $\beta=0.0,~0.5,~0.7$, and 0.9. In the case where the accretor is a degenerate star, however, mass accretion proceeds at the Eddington accretion rate. When $\beta=0.0$, masses are not lost from a binary system with RLO, so that this case is called conservative mass transfer.


\subsection{Common envelope\label{subsec:ce}}
Undergoing a CE phase, binary systems merge or become tighter. Thus, CE plays important roles in determining binary fates. We employ a criterion that the core of the donor star merges with the companion after a CE phase if $R'_1+R'_2>a_\mathrm{f}$, where $R'_1$, $R'_2$, and $a_\mathrm{f}$ are the radius of the donor star, the radius of the companion, and the separation after the CE phase, respectively \citep{belczynski08}. Although CE plays important roles in the binary evolution, the separation after a CE phase has large uncertainties. The reason for this is that 3D hydrodynamical simulations cannot yet reveal the ejection process during a CE phase. Furthermore, an observational restriction has not yet been imposed.

In our calculation, we use the following formulation to evaluate the separation of binaries after a CE phase,
\begin{align}\label{eq:ce}
   \alpha_\ce\left(\frac{GM_\mathrm{c,1}M_2}{2a_\mathrm{f}}-\frac{GM_1M_2}{2a_\mathrm{i}}\right)=\frac{GM_1M_\mathrm{env,1}}{\lambda R_1},
\end{align}
where $M_1$, $M_\mathrm{c,1}$, $M_\mathrm{env,1}$, $R_1$, $M_2$, $a_\mathrm{i}$, and $a_\mathrm{f}$ are the mass of the donor star, the core mass of the donor star, the envelope mass of the donor star, the radius of the donor star, the mass of the companion star, the separation before the CE phase, and the separation after the CE phase, respectively \citep{webbink84}. $\alpha_\ce$ is the efficiency with which the released orbital energy ejects the envelope of the companion. In general, the value of $\alpha_\ce$ is less than unity, but if additional energy such as recombination energy or nuclear energy helps ejection, the value of $\alpha_\ce$ can become larger than unity \citep{podsiadlowski10,ivanova13}. In this study, we set the value of $\alpha_\ce$ to 1.0. $\lambda$ is the binding energy parameter and its value changes with the evolution \citep{dewi00,xu10} and we use the fitting formulae of \citet{xu10,xu10e}. They presented their fitting formulae of $\lambda$ for the following two cases: the fraction of the internal energy contributing to ejection $\alpha_\mathrm{th}$ = 0, and 1. We use the average of these. Additionally, we conduct calculations for three fixed binding energy parameters $\lambda=0.1,~0.5$, and 1.0 for comparison with previous works.

When the donor star is in the HG on the Hertzsprung--Russell diagram, its core-envelope structure is not clear. Therefore, it is considered that the orbital energy is used to eject not only the envelope but the entire star, so that the ejection is hard to complete, and thus the system results in coalescence \citep{belczynski07}. In this work, we consider following two cases. One is to ignore this fact and estimate the separation using equation (\ref{eq:ce}) for donor stars in the HG as well as giant stars that have a clear core-envelope structure \citep[submodel A in][]{dominik12}. The other is to assume that if a binary system enters a CE phase when the donor star is in the HG, it always coalesces \citep[submodel B, or the pessimistic approach in][]{dominik12}. In this paper, we call the latter a ``pessimistic" CE. The same processing is done to donor stars without hydrogen envelopes, namely helium stars in the HG (helium stars that finished core helium burning but have not yet started helium shell burning).


\subsection{Supernovae\label{subsec:sn}}
\begin{table}[t]
 	\caption{The 1D Standard Deviations, $\sigma$, of Maxwellian distribution for each type of SNe.\label{table:sigma}}
 	\begin{center}
      \begin{tabular}{cc}\hline\hline
         SNe Type & $\sigma~(\mathrm{km}~\psec)$\\\hline
         Canonical CCSNe & 265\\
         US CCSNe & 30\\
         Canonical ECSNe & 30\\
         US ECSNe & 30\\\hline
      \end{tabular}
   \end{center}
\end{table}
SN explosion can also change the orbital parameters of systems drastically. We consider the following four types of SNe: canonical core-collapse SNe (CCSNe), canonical ECSNe, US CCSN, and US ECSN. For CCSNe, to calculate the remnant mass from the CO core mass, we adopt the ``rapid'' SN mechanism in \citet{fryer12}. The characteristic of this mechanism is that an accretion onto PNS during explosion is forbidden. 

In this study, we set the value of the Chandrasekhar mass to $1.37~\msun$ and assume that the baryonic mass of an NS formed through ECSNe is the same as the Chandrasekhar mass. In our code, a single star with its ZAMS mass 7.56--8.17 $\msun$ evolves to have the initial carbon--oxygen (CO) core mass $1.34\le M_\mathrm{CO}/\msun\le 1.37$. It may increase its CO core mass by the helium shell burning, and finally cause ECSN. In binary systems, the ZAMS mass range of ECSN progenitors becomes wider due to the mass-loss and gain \citep[e.g.,][]{podsiadlowski04}. We assume that a CO core whose mass reaches the Chandrasekhar mass causes the ECSN explosion.

We define USSNe as having the helium envelope mass less than $0.2~\msun$ just before SN \citep{tauris15}. Thus, canonical CCSN (ECSN) means an SN with its progenitor having the helium envelope mass $M_\mathrm{He,env}$ greater than $0.2~\msun$ just before the explosion. In addition, we define iPTF 14gqr like USSNe as USSNe ejecting the mass of $M_\mathrm{ej}=0.15$--$0.30~\msun$ containing the helium envelope of $M_\mathrm{He,env}=0.003$--$0.013~\msun$.

A pulsar kick imparted to PNS changes orbital parameters. We adopt a bimodal kick velocity distribution \citep{katz75,verbunt17}. \citet{hobbs05} show that kick velocities can be described by 1D Maxwellian distribution with its standard deviation $\sigma=265~\mathrm{km}~\psec$ from the observation of proper motions of pulsars. Therefore, we simply assume that kick velocities of all types of SN follow the Maxwellian distribution regardless of whether the primary or secondary explodes. The standard deviation for canonical CCSNe is assumed to be $\sigma=265~\mathrm{km}~\psec$ \citep{hobbs05}. Following \citet{pfahl02} and \citet{podsiadlowski04}, we set the standard deviation for ECSNe and USSNe to $\sigma=30~\mathrm{km}~\psec$. The assumed standard deviations for each type of SNe are in Table \ref{table:sigma}. 


\section{Results\label{sec:results}}
\begin{table*}[tb]
 	\caption{The obtained formation rates and merger rates of DNS systems and the rate of USSNe.\label{table:results}}
 	\begin{center}
      \begin{tabular}{ccccccccccc}\hline\hline
         Rank & Model & $\beta$ & ``pessimistic "CE & $R_{f,\mathrm{DNS}}$ & $R_{m,\mathrm{DNS}}$ & USSNe & $\log\tilde{\Lambda}$\\
         &&&&($\pgal~\pmyr$)&($\pgal~\pmyr$)&($\pgal~\pmyr$)&\\\hline
         1 &      5.0 & 0.5 & 0 &   12.71 &    4.94 &  510.88 (16.31) & $ 0$\\
         2 &      5.1 & 0.5 & 1 &   12.50 &    4.82 &  488.80 (14.29) & $ -0.13$\\
         3 &      7.1 & 0.7 & 1 &   17.50 &    3.00 &  755.77 (37.94) & $ -1.52$\\
         4 &      7.0 & 0.7 & 0 &   18.76 &    3.84 &  785.90 (40.61) & $ -1.56$\\
         5 &      9.0 & 0.9 & 0 &   21.51 &    7.49 &  280.67 (12.84) & $ -1.94$\\
         6 &      9.1 & 0.9 & 1 &   17.54 &    5.40 &  268.87 (11.86) & $ -1.99$\\
         7 &      0.0 & 0.0 & 0 &    8.51 &    4.49 &  543.11 (20.31) & $ -3.72$\\
         8 &      0.1 & 0.0 & 1 &    7.96 &    4.04 &  539.12 (20.13) & $ -3.75$\\\hline
      \end{tabular}
   \end{center}
   \tablecomments{In column 7, the rates given in parentheses are those of iPTF 14gqr like USSNe.}
\end{table*}
In this paper, we denote our models as 10X.Y. Here, X=$\beta$, and Y=1, or 0 represents whether or not ``pessimistic'' CE is considered. For example, the model 5.0 is the model for $\beta=0.5$, and not considering ``pessimistic'' CE. Table \ref{table:results} shows the results of the eight models in which $\lambda_\nj$ is used, and we arrange them in descending order of likelihood. The first column is the rank of the likelihood. The second column is the model name. The third column is $\beta$ for each model. The fourth column represents whether ``pessimistic'' CE is considered or not. In the case for 0, ``pessimistic'' CE is not considered. The fifth column is the formation rate of DNS systems. The sixth column is the merger rate of DNS systems. 
The seventh column is the rate of USSNe (the rates of iPTF 14gqr like USSNe are given in brackets). The eighth column, $\tilde{\Lambda}$ is the ratio of each model's likelihood to the largest likelihood. The full set of our results, including models of fixed $\lambda$, is in the Appendix \ref{append1}.

We obtain that model 5.0 is the maximum likelihood model for the observed 15 DNS systems. In this model, iPTF 14gqr like USSNe occur at the rate of $R_\mathrm{14gqr}=16.31~\pgal~\pmyr$. Hereafter, we estimate the detection rate of iPTF 14gqr like USSNe by some optical transient surveys using model 5.0. To estimate the detection rate, we utilize the following three assumptions. First, SNe with the same ejecta and helium envelope mass as iPTF 14gqr reproduce the light curve of iPTF 14gqr. To obtain the information about the envelope, we must observe the shock cooling emission phase. Therefore, we set the lowest absolute magnitude of iPTF 14gqr like USSNe as $-16.6$ mag in the {\it R} band \citep[See Fig. 2 in][]{de18}. This is the second assumption. Finally, we assume that an object must be observed every two days. 

Under these three assumptions, the maximum distance $D$ that iPTF 14gqr like USSNe can be observed is
\begin{align}
   D=&10~[\mathrm{pc}]~\times10^{(m+16.6)/5}\notag\\
   =&10^{m/5-1.68}~[\mathrm{Mpc}],
\end{align}
where $m$ denotes the limiting magnitude of a survey. Hence, the detection rate of iPTF 14gqr like USSNe is
\begin{align}\label{eq:detection}
   &\frac{4\pi}{3}D^3~[\mathrm{Mpc}^3]~\times\rho_\mathrm{gal}~[\mathrm{galaxies~Mpc}^{-3}]\notag\\
   &~~~~~~~~~~~~~~~~~~~\times~R_\mathrm{14gqr}~[\pgal~\mathrm{Myr^{-1}}]\notag\\
   &~~~~~~~~~~~~~~~~~~~\times\frac{A~[\sdeg]}{41253~[\sdeg]}\notag\\
   =&1.8\times10^{-13}\times10^{3m/5}\times\frac{A~[\sdeg]}{1000~[\sdeg]}~[\pyr],
\end{align}
where $\rho_\mathrm{gal}=0.0116~\mathrm{galaxies~Mpc}^{-3}$ and $A$ are the Milky Way--like galaxy density \citep{dominik13}, and a survey area with a cadence of $<2~\mathrm{days}$, respectively. In the following, we assume that the duty cycle is 100\%. 


\subsection{intermediate Palomar Transient Factory\label{subsec:iptf}}
iPTF \citep{law09,rau09} uses CFHT 12k mosaic camera on the 48 inch (1.2 m) Samuel Oschin Schmidt Telescope at the Palomar Observatory (P48). The field of view of this camera is $7.8~\sdeg$. The standard exposure time is 60 s and the limiting magnitude is $m=21~\mathrm{mag}$ in the {\it R} band. The iPTF spends 41\% of its total time for surveying with a cadence of 5 days. 40\% is used to execute an experiment with a cadence of 1 minute--3 days. This experiment is called the dynamical cadence (DyC). The remaining 11\% and 8\% are monitoring the Orion star forming region and all-sky survey with a narrowband filter, respectively. 

iPTF 14gqr like USSNe can be detected only during DyC. We assume that a cadence during DyC is always 1 day. The iPTF can cover 1000 $\sdeg$ per one night, so that the survey area with a cadence $<2~\mathrm{days}$ is $A=1000~\sdeg$. Therefore, we evaluate the detection rate 0.3$~\pyr$ from equation (\ref{eq:detection}). Note that a factor 0.4 has to be multiplied to obtain this detection rate because the survey time of DyC is 40\% of the total time. The iPTF ran from 2013 January 1 to 2017 March 2 (4.17 yr), so that $1$ iPTF 14gqr like USSN can be detected totally. Furthermore, if we consider not only iPTF 14gqr like USSNe but also other USSNe, for example, a USSN which has more or less helium envelope than the iPTF 14gqr, the detection rate becomes larger. Thus, there may be other detections of USSNe by iPTF.


\subsection{Zwicky Transient Facility\label{subsec:ztf}}
ZTF \citep{bellm17} uses a new camera with the 47 $\sdeg$ field of view on the P48. The standard exposure time is 30 s and the limiting magnitude is $m=20.5~\mathrm{mag}$ in the {\it R} band. ZTF can scan the whole sky in one night, so the survey area is $A\sim30,000~\sdeg$. From equation (\ref{eq:detection}), we find that ZTF can detect USSNe at a rate of $10~\pyr$.


\subsection{Large Synoptic Survey Telescope\label{subsec:lsst}}
LSST \citep{ivezic08} uses a camera with the 9.6 $\sdeg$ field of view on an 8.4 m telescope, and executes surveys across the southern hemisphere. The cadence is $\sim 4$ days during 90\% of the total survey time. The remaining 10\% is spent for surveying with a very shorter cadence ($\sim1$ minute) and the limiting magnitude is $\sim26.5$ mag in the {\it R} band. This experiment with a very short cadence is called deep-drilling fields. Only during this experiment can we detect iPTF 14gqr like USSNe. We assume that the survey area $A$ is equal to the field of view of the camera 9.6 $\sdeg$. From the above, we find that LSST can detect USSNe at a rate of 1 $\pyr$.

Finally, we estimate that general USSNe with the helium envelope mass $M_\mathrm{He,env}$ less than $0.2~\msun$ occur at the rate of $5.1\times10^2~\pgal~\pmyr$ for the most favored model. Using the fact that the Galactic total SNe rate is $4.6^{+7.4}_{-2.7}~\mathrm{century}^{-1}$ \citep[][includes SNe Ia]{adams13}, we can calculate that the ratio of USSNe to total SNe is 0.004--0.03. \citet{tauris13} estimate that the ratio of USSNe to total SNe is 0.001--0.01. Our obtained value is roughly consistent with this.

\section{Discussion} \label{sec:discussion}
\begin{figure*}[t]
   \begin{minipage}{0.5\hsize}
      \begin{center}
         \includegraphics[width=\columnwidth,bb=0 0 1209 995]{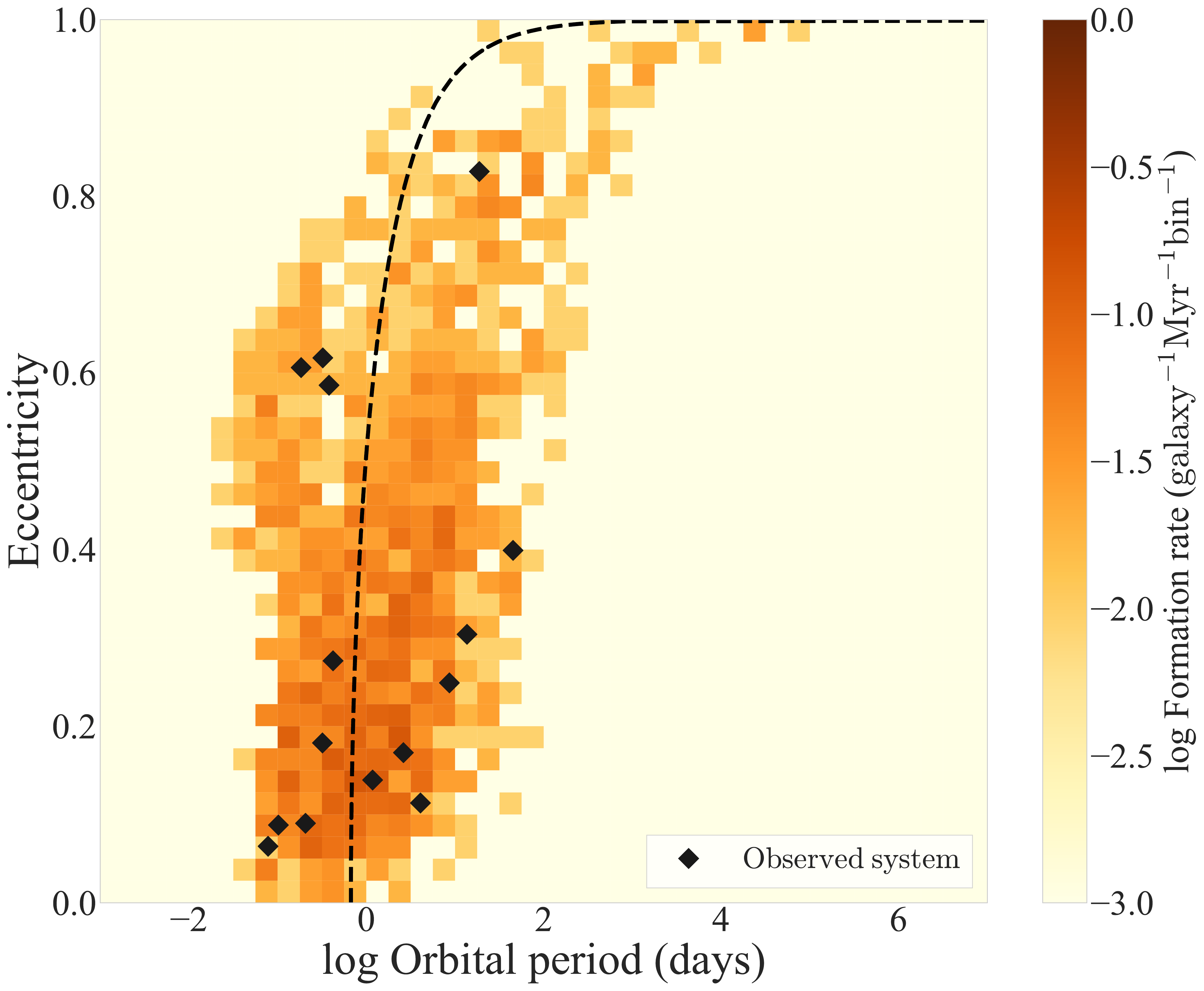}
      \end{center}
   \end{minipage}
   \begin{minipage}{0.5\hsize}
      \begin{center}
         \includegraphics[width=\columnwidth,bb=0 0 1209 995]{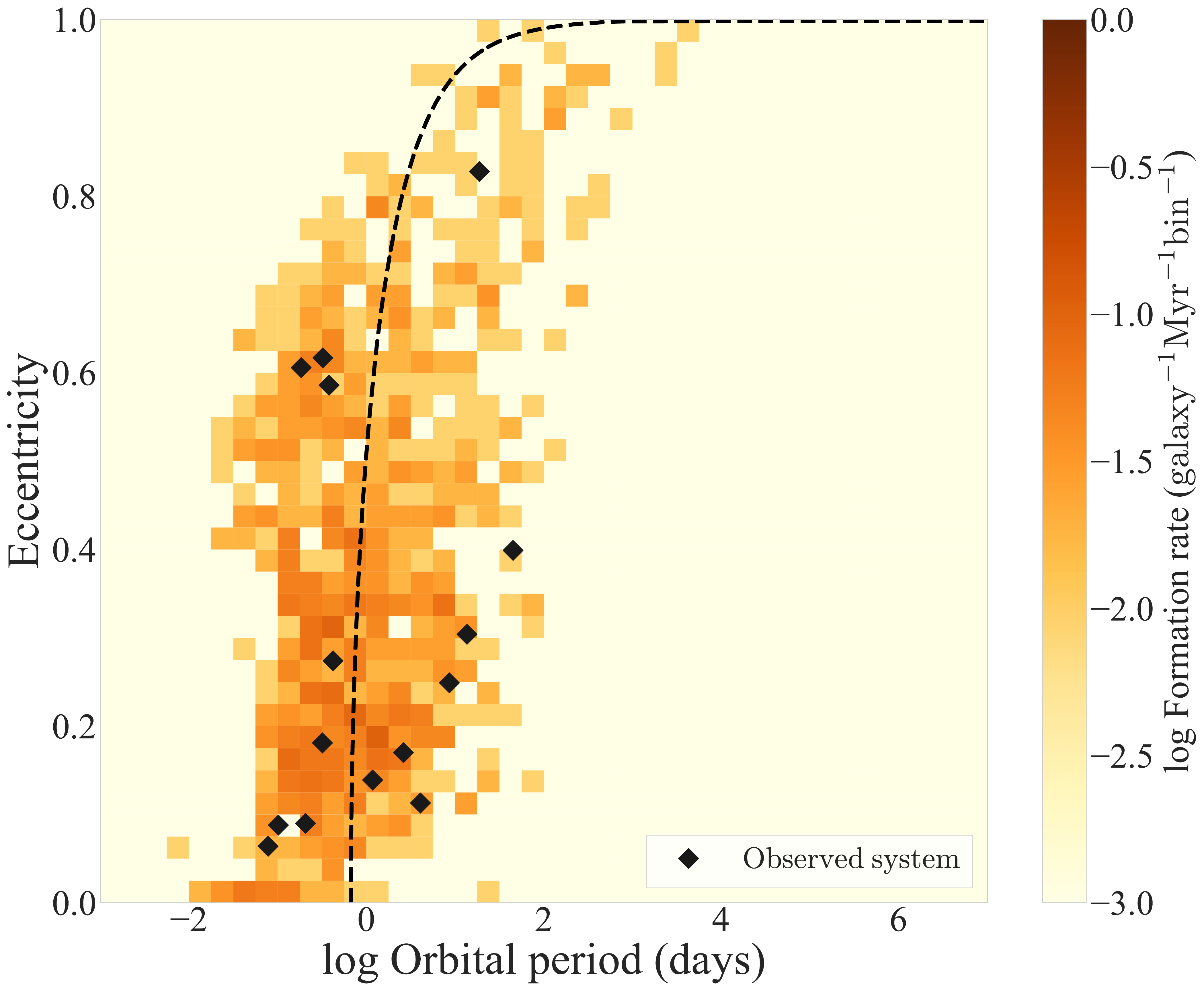}
      \end{center}
   \end{minipage}
      \caption{Formation rate distribution of the most favored model 5.0 (left; $\beta=0.5$, and ``pessimistic'' CE is not considered) and its variation 0.0 (top right; $\beta=0.0$). The dashed line shows the orbital parameters of DNS systems that will merge within a Hubble time. A bin size is $\Delta(\log P)=0.25$ and $\Delta e=0.025$, and there are 1600 bins in total.\label{fig:prob}}
\end{figure*}


\subsection{Merger rate} \label{subsec:rate}
\begin{table}[b]
   \caption{DNS Merger Rates Derived by Many Authors.\label{table:dns_rate}}
   \begin{center}
      \begin{tabular}{ccc}\hline\hline
      Ref. & DNS Merger Rate & Method \\
      & ($\pgal~\pmyr$) &\\\hline
      1 & 4.8\tablenotemark{a} & Population synthesis\\
      2 & 3 & Population synthesis\\
      2 & 40\tablenotemark{a,b} & Population synthesis\\
      3 & 4 & Population synthesis\\
      4 & 24.04 & Population synthesis\\
      5 & $21^{+28}_{-14}$ & Galactic DNS observation\\
      6 & $154^{+320}_{-122}$\tablenotemark{a} & GW detection\\\hline
      \end{tabular}
   \end{center}
   \tablenotetext{a}{The local merger rate density ($\pgpcc~\pyr$) is provided in the paper. We convert this into the galactic merger rate simply using $\rho_\mathrm{gal}=0.01$.}
   \tablenotetext{b}{The authors' upper limit.}
   \tablerefs{(1) \citet{chruslinska18}; (2) \citet{kruckow18}; (3) \citet{shao18}; (4) \citet{vigna18}; (5) \citet{kim15}; (6) \citet{abbott17c}}
\end{table}

Our calculated merger rate of DNS systems is 4.94 $\pgal~\pmyr$, and approximately equal to 50 $\pgpcc~\pyr$. This value is roughly consistent with recent population synthesis studies. \citet{vigna18} use population synthesis code COMPAS \citep{stevenson17} and assume that the kick distribution is bimodal and case BB mass transfer is always dynamically stable. They estimate a Galactic merger rate of $24.04~\pgal~\pmyr$. \citet{shao18} use the BSE code \citep{hurley02} and predict a merger rate of $4~\pgal~\pmyr$. Their mass-loss efficiency $\beta$ depends on the angular velocity of the accretor, and highly nonconservative mass transfer is assumed. \citet{kruckow18} use {\sc ComBinE} \citep[upgraded from a code developed by][]{voss03}, their merger rate is $3~\pgal~\pmyr$, and the upper limit is $400~\pgpcc~\pyr$ in the local universe. \citet{chruslinska18} use {\sc StarTrack} \citep{belczynski02,belczynski08} and estimate the merger rate of $48~\pgpcc~\pyr$. As can be seen from Table \ref{table:results} and other authors' merger rates, there is little difference between population synthesis calculations. This suggests that it is difficult to constrain binary physics from DNS merger rate.

On the other hand, \citet{kim15} obtain the Galactic DNS merger rate of $21^{+28}_{-14}~\pgal~\pmyr$ based on the observed Galactic DNS systems. This is also consistent with our results and recent population synthesis studies. \citet{abbott17c} estimate the merger rate of $1540^{+3200}_{-1220}~\pgpcc~\pyr$ using the only detected DNS merger event, GW1708017. This is higher than the above values. If the observational period becomes longer and other events are detected in the future, this estimation may vary.

DNS merger rates derived by many authors are given in Table \ref{table:dns_rate}. We roughly assume that all DNS merger events involve one USSN and that all of these USSNe reproduce light curves that are the same as that of iPTF 14gqr; thus we can estimate the detection rate of USSNe using Equation (\ref{eq:detection}). The obtained detection rates by iPTF, ZTF, and LSST are 0.05--9, 2--300, and 0.2--30 $\pyr$, respectively. Note that USSNe may occur when DNS systems that do not merge within a Hubble time and NS--BH binaries are produced \citep{kruckow18}. In our most favored model 5.0, the ratio of merging DNS systems which produced via a canonical channel that the secondary causes USSN to all DNS systems produced via canonical channel is 40\%. Thus, these results may be underestimation.


\subsection{Binary physics} \label{subsec:binary_phys}
\subsubsection{mass-loss efficiency $\beta$} \label{subsubsec:beta}
\begin{table*}[t]
 	\caption{Ratios of DNS systems produced via each channel for the model 5.0 ($\beta=0.5$).\label{table:channel_5.0}}
 	\begin{center}
      \begin{tabular}{ccc}\hline\hline
         Channel & Ratio to & Ratio to \\
         & all DNSs (\%) & DNSs that Merge within a Hubble Time (\%)\\\hline
         No CE & ~5.5 & ~0.4\\
         Single-core channel & 89.2 & 91.4\\
         Double-core channel & ~5.2 & ~7.5\\
         2 CEs & ~0.0 & ~0.0\\\hline
      \end{tabular}
   \end{center}
\end{table*}
\begin{table*}[t]
 	\caption{Same as Table \ref{table:channel_5.0}, but for the model 7.0 ($\beta=0.7$).\label{table:channel_7.0}}
 	\begin{center}
      \begin{tabular}{ccc}\hline\hline
         Channel & Ratio to & Ratio to \\
         & all DNSs (\%) & DNSs that Merge within a Hubble Time (\%)\\\hline
         No CE & 50.3 & ~6.8\\
         Single-core channel & 40.9 & 65.4\\
         Double-core channel & ~2.8 & ~6.4\\
         2 CEs & ~0.6 & ~1.7\\\hline
      \end{tabular}
   \end{center}
\end{table*}
\begin{table*}[t]
 	\caption{Same as Table \ref{table:channel_5.0}, but for the model 9.0 ($\beta=0.9$).\label{table:channel_9.0}}
 	\begin{center}
      \begin{tabular}{ccc}\hline\hline
         Channel & Ratio to & Ratio to \\
         & all DNSs (\%) & DNSs that Merge within a Hubble Time (\%)\\\hline
         No CE & 22.7 & ~1.1\\
         Single-core channel & 15.4 & 10.4\\
         Double-core channel & 38.2 & 67.8\\
         2 CEs & 22.5 & 17.5\\\hline
      \end{tabular}
   \end{center}
\end{table*}
To explore the effect of the value of mass-loss efficiency $\beta$ during RLO on the binary evolution and the ultimate fates, we consider 4 constant values 0.0, 0.5, 0.7, and 0.9 for $\beta$. As can be seen in Table \ref{table:results}, the effect of mass-loss efficiency during RLO on the DNS merger rate is little. 

From Table \ref{table:results}, it is suggested that $\beta=0.5$, that is, half nonconservative mass transfer is preferable to explain the orbital parameters of the observed Galactic DNS systems. The larger the value of $\beta$ is, the more masses are lost. Then the orbital separation can become longer. Therefore, in the case where $\beta$ is small, binary systems with long orbital periods are hard to form. We show in Figure \ref{fig:prob} the formation rate distribution of the most favored model 5.0 (left; $\beta=0.5$, and ``pessimistic'' CE is not considered) and its variation model 0.0 (right; $\beta=0.0$). There are several systems with $e=0$. In our calculation, some systems cause second explosion during a CE phase. We assume that the eccentricity gets to zero after the CE phase, so that these systems can become circular orbit DNS systems. Compared with the left ($\beta=0.5$), the right ($\beta=0.0$; conservative mass transfer) hardly produces binary systems with longer orbital periods such as PSR J1930--1852 \citep{swiggum15}.


\subsubsection{Common envelope phase} \label{subsubsec:ce2}
To treat the CE process, we use $\lambda_\nj$ \citep{xu10,xu10e}, and $\alpha_\ce=1.0$. We also use fixed $\lambda$ for comparison with previous works. Our DNS merger rates of model 5.0 ($\beta=0.5$, ``pessimistic'' CE is not considered) and its variations model on $\lambda$ ($\lambda=0.1,~0.5$,and 1.0) are 4.94, 10.85, 51.50, and 66.09 $\pgal~\pmyr$, respectively (see Table \ref{table:results_full}). The merger rate varies by an order of magnitude due to the change of $\lambda$. Especially, when $\lambda$ is as large as 0.5 or 1.0, the binary systems are likely to be able to avoid coalescence during CE, then the merger rate gets higher.

We consider two cases where ``pessimistic'' CE is considered or not. ``Pessimistic'' CE means that if a CE phase starts when the donor star is in the HG, the binary system will always coalesce. Therefore, close binary systems are harder to form if ``pessimistic'' CE is not considered. Our DNS merger rates of model 5.0 and 5.1 (``pessimistic'' CE is considered) are 4.94 and 4.82 $\pgal~\pmyr$, respectively. Although the formation rate and thus merger rate should become smaller when ``pessimistic'' CE is considered, the effect seems to be little.

Recently, \citet{kruckow18} argued that applying calculated $\lambda$-values to stellar grids based on a different stellar evolution code may lead to serious inconsistencies, compared to the self-consistent manner. Thus, some of our obtained merger rates may change if $\lambda$-values are calculated in a self-consistent manner.


\subsubsection{Electron-capture supernovae} \label{subsubsec:ecsne}
\begin{figure*}[t]
   \begin{minipage}{0.33\hsize}
      \begin{center}
         \includegraphics[width=\columnwidth,bb=0 0 1001 995]{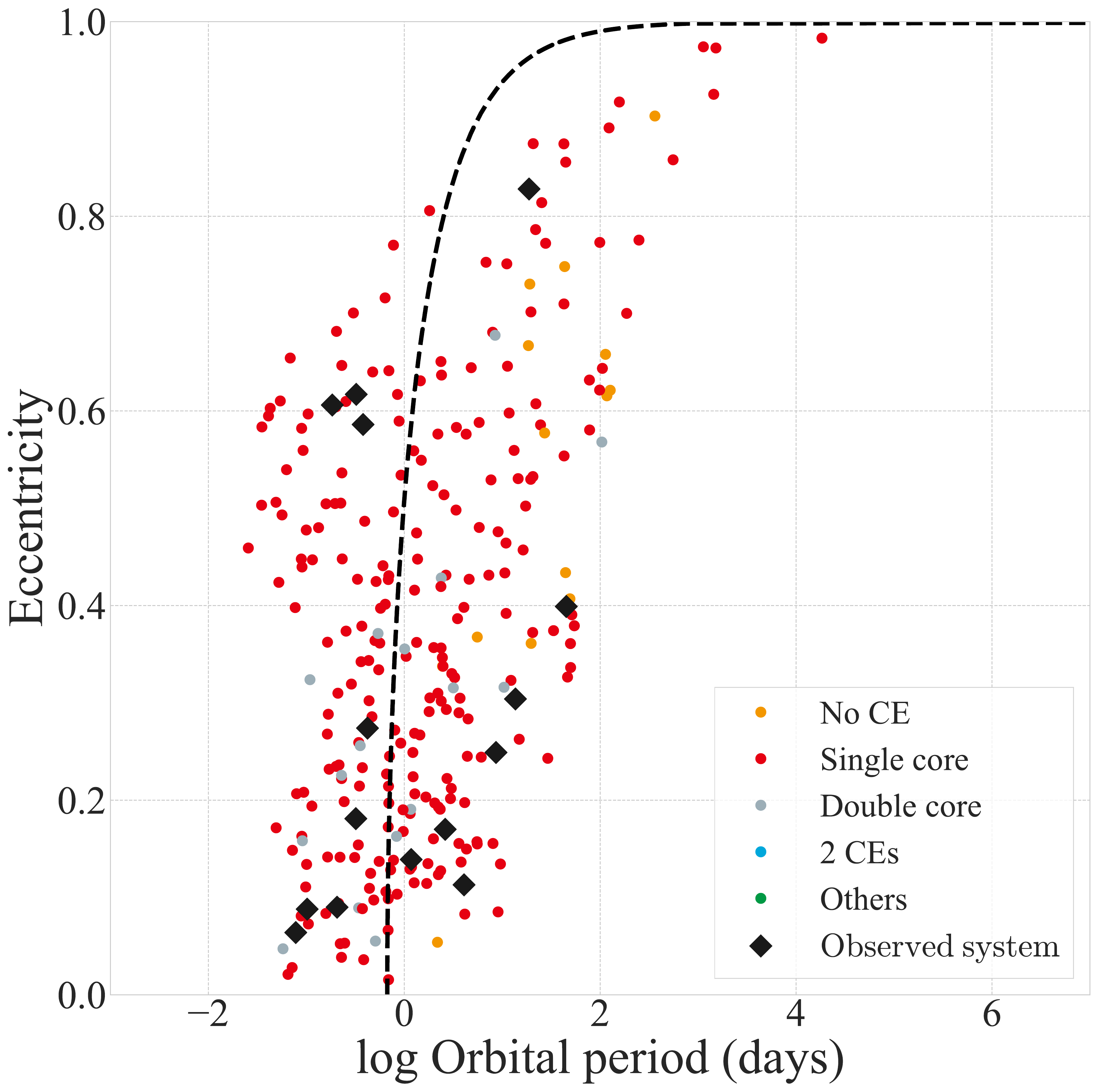}
      \end{center}
   \end{minipage}
   \begin{minipage}{0.33\hsize}
      \begin{center}
         \includegraphics[width=\columnwidth,bb=0 0 1001 995]{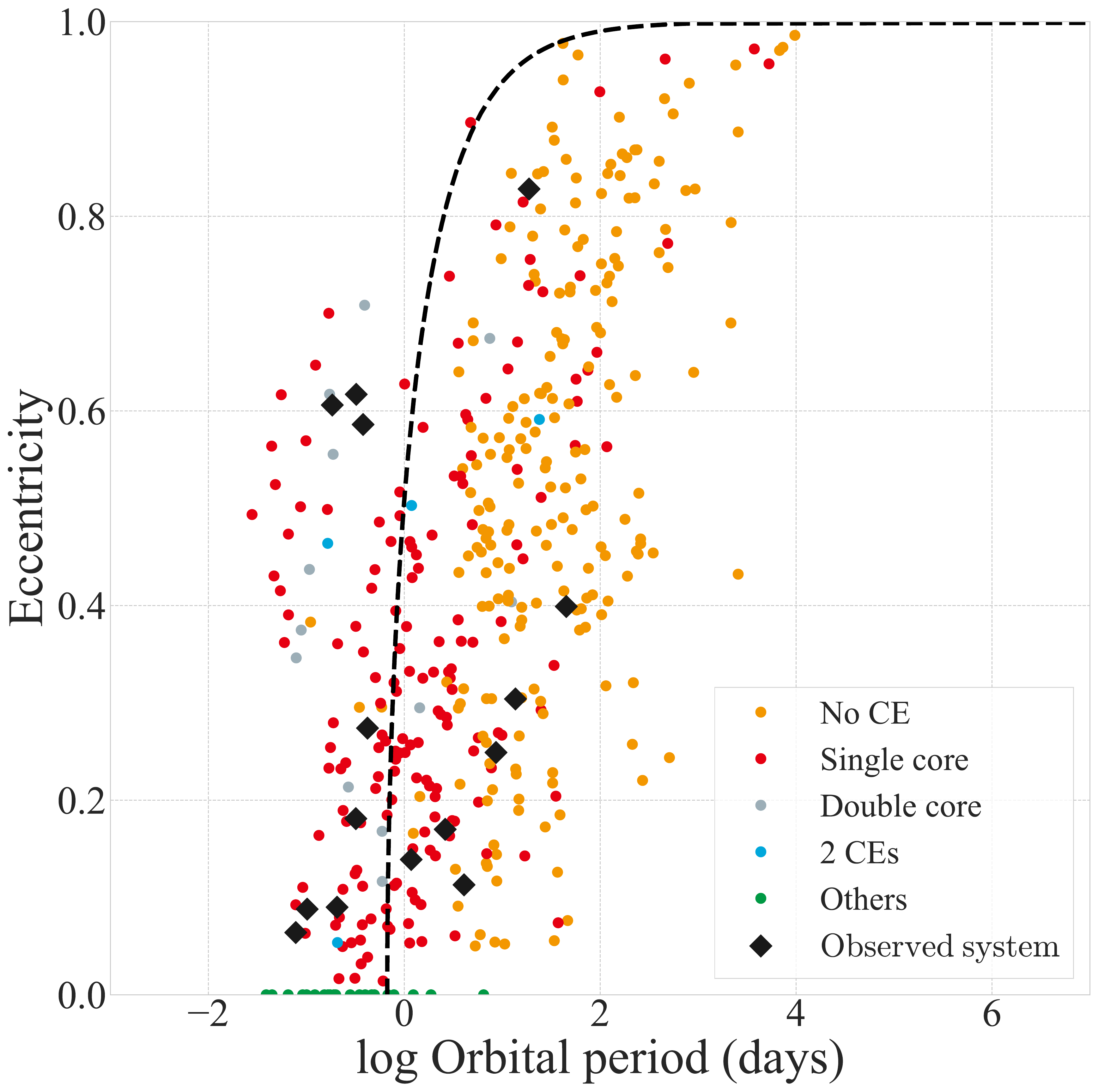}
      \end{center}
   \end{minipage}
   \begin{minipage}{0.33\hsize}
      \begin{center}
         \includegraphics[width=\columnwidth,bb=0 0 1001 995]{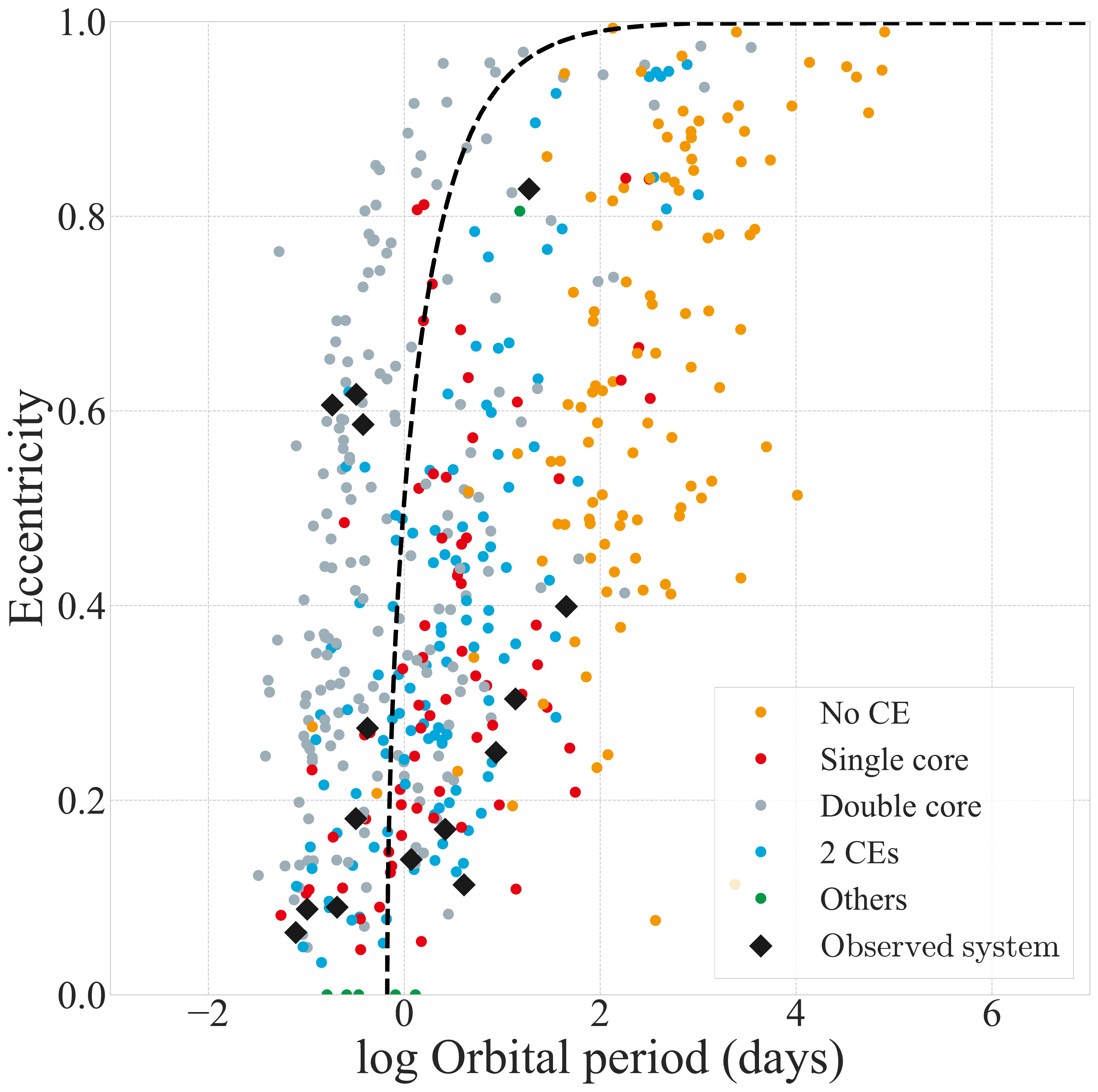}
      \end{center}
   \end{minipage}
   \caption{The orbital parameters of produced DNS systems at birth for the model 7.0 (left; $\beta=0.7$, and ``pessimistic CE'' is not considered) and its variation 9.0 (right; $\beta=0.9$, and ``pessimistic CE'' is not considered). To make it easier to see tendencies, we show the results of only $2\times10^5$ binaries. The dashed line shows the orbital parameters of DNS systems that will merge within a Hubble time. The colors denote the formation channel. Orange represents DNS systems that experience no CE phase. Red and gray represent single-core and double-core channels, respectively. DNS systems that experience CE phase twice are blue. Finally, green shows other channels. For example, systems that cause a second SN explosion during the CE phase are classified as other channels and denoted by green.\label{fig:channel}}
\end{figure*}
From the  distribution of the mass and orbital parameters of observed Galactic DNS systems, \citet{beniamini16} show that the kick velocity of second explosion may be small. In our study, US CCSNe, canonical ECSN, and US ECSN can achieve the small pulsar kick (see Table \ref{table:sigma}). In our most favored model, the fraction of DNS systems with second NS formed via canonical ECSNe or US ECSNe to all DNS systems is 5.0\%. 88.4 and the remaining 6.6\% of all DNS systems have a second NS formed via US CCSNe and canonical CCSNe, respectively. Thus, it is suggested from our results that the inferred small kicks of observed DNS systems mainly arise from US CCSNe, not ECSNe.

\citet{shao18} remark that the fraction of DNS systems that the second NS is formed via ECSNe among all DNS systems is 0.74 for their default model (the helium core mass range of the single ECSN progenitor is $1.83$--$2.75~\msun$ and the 1D standard deviation of the kick velocity of ECSNe is $\sigma=40~\mathrm{km}~\mathrm{s^{-1}}$). This value seems to be inconsistent with our results: 0.05. However, they do not consider USSNe and they also remark that their range of $1.83$--$2.75~\msun$ is wider than previous studies and this wideness may include contributions of USSNe. Their suggestion is consistent with our results. 


\subsubsection{DNS systems with high eccentricity and short period} \label{subsubsec:subpopiii}
\citet{andrews19} claim that the DNS systems with $P\sim 0.3~\mathrm{days}$, and $e\sim 0.6$ (same as Hulse--Taylor DNS, and we call these population as subpopulation (iii) as with \citet{andrews19}) can be produced by primordial binary evolution only when helium star mass is $M_\mathrm{He}\simeq 3.2~\msun$, and kick magnitude is $w_\mathrm{k}\lesssim 25~\mathrm{km~s^{-1}}$. They say that it is unlikely that the SN explosion of helium star $M_\mathrm{He}\simeq 3.2~\msun$ unlikely receives a small kick such as $w_\mathrm{k}\lesssim 25~\mathrm{km~s^{-1}}$, hence the subpopulation (iii) DNSs cannot be formed by primordial binary evolution.

On the other hand, as seen in Figure \ref{fig:prob}, subpopulation (iii) DNSs are formed by binary evolution in our calculation. For example, when helium star mass is $M_\mathrm{He}=2.7~\msun$ (CO core mass is $M_\mathrm{CO}=2.59~\msun$, helium envelope is $M_\mathrm{He,env}=0.11~\msun$, so that subsequently causes USSN), first NS mass is $M_\mathrm{NS,1}=1.25~\msun$, second NS mass is $M_\mathrm{NS,2}=1.13~\msun$, pre-second-SN separation is $a_0=1.0~\mathrm{R}_\odot$, and kick magnitude is $w_\mathrm{k}=40~\mathrm{km~s^{-1}}$, the subpopulation (iii) DNSs can be formed, and this is not the rare case. In our calculation, exploding helium stars, which are much lighter than $2.7~\msun$, cannot create eccentric DNS systems due to the small mass ejection. Note that the masses of NSs are gravitational masses.

The difference from \citet{andrews19} is the NS mass. Whereas they consider only NSs with $1.4~\msun$, we use the `rapid' SN mechanism from \citet{fryer12}, and this mechanism has a tendency to create somewhat light NS ($\lesssim 1.3~\msun$). Thus, our calculation can produce the subpopulation (iii) DNSs.


\subsection{Formation channel} \label{subsec:channel}
Various formation channels leading to DNS systems have been proposed by many sutudies. In our calculation, there are two main channels. One is the canonical channel \citep[see Figure 1 in ][]{tauris17}. After first explosion, a binary system becomes HMXB and undergoes the CE phase, then the secondary, stripped by the case BB RLO, explodes. In this section, we call this channel the single-core channel. The other is that the initial mass ratio is approximately equal to unity and binary systems experience the double-core CE phase \citep{brown95}. The double-core CE means that both stars with clear core-envelope structures enter the CE phase before the first SN \citep[See Figure 6 in][]{vigna18}. 

We show in Figure \ref{fig:channel} the orbital parameters of DNS systems for the model 5.0 (left), 7.0 (middle) and 9.0 (right) and the colors denote the formation channel. The ratios of DNS systems produced via each channel for model 5.0 and 7.0 are shown in Table \ref{table:channel_5.0}, \ref{table:channel_7.0}. From the left panel of Figure \ref{fig:channel} and Table \ref{table:channel_5.0}, we find that almost all of DNS systems (91.4\%) that will merge within a Hubble time are produced via single-core channel whereas 7.5\% are produced via double-core channels. For model 7.0, 65.4\% (6.4\%) of merging DNS systems are formed via single-core (double-core) channels. The number of merging DNS systems formed via single-core channels is larger than that of merging DNS systems formed via double-cor channels by an order of magnitude for not extremely high $\beta$.

Table \ref{table:channel_9.0} is same as Table \ref{table:channel_5.0} but for the model 9.0, which is a variation of the model 5.0. From the right panel of Figure \ref{fig:channel} and Table \ref{table:channel_9.0}, we find that the ratio of DNS systems formed via the double-core channel for the model 9.0 is 67.8\% and is greater than that of the model 7.0. If the mass transfer is not extremely nonconservative (for example, $\beta=0.5$, and 0.7), the evolution of the accretor gets slower due to the mass accretion and then the period during which the accretor is in the main-sequence becomes longer. Thus, it is hard for the double-core channel to occur. On the other hand, in the case of extremely higher $\beta$ , such as 0.9, the effect of rejuvenation is smaller. Hence, the double-core channel occurs to some degree. We may be able to constrain the value of $\beta$ from observations of intermediate products of each channel.


\section{Summary} \label{sec:summary}
We perform rapid population synthesis calculation to estimate the detection rate of iPTF 14gqr like USSNe by some optical transient surveys. We consider four values for the mass-loss efficiency, $\beta=0.0,~0.5,~0.7$, and 0.9, and for two cases: if ``pessimistic'' CE is considered or not. For each model, we calculate a probability distribution of the orbital parameters of produced DNS systems and then evaluate the likelihood using the observed 15 Galactic DNS systems as samples. Our most favored model is $\beta=0.5$ and ``pessimistic'' CE is not considered. Our obtained DNS merger rate is 4.94 $\pgal~\pmyr$. This is roughly consistent with other studies except for the estimation by the GW detection.

We find that iPTF and next-generation optical synoptic surveys ZTF and LSST can observe iPTF 14gqr like USSNe at the rate of 0.3, 10, and 1 $\pyr$, respectively. The iPTF ran for 4.17 years, so that 1 iPTF 14gqr like USSNe could be detected.  With the next-generation survey, ZTF, we may find the population of USSNe and discover some anomalous USSNe. Additionally, we roughly assume that all DNS merger events involve one USSN and all of these USSNe reproduce light curves that are the same as that of iPTF 14gqr, then we can estimate the detection rate of USSNe from DNS merger rates derived by many authors. Using this method, we find that iPTF, ZTF, and LSST can detect USSNe at the rate of 0.05--9, 2--300, and 0.2--30 $\pyr$, respectively.

We also find that 90\% of DNS systems have a second NS that is formed via US CCSNe, so that US CCSNe rather than ECSNe play important roles to produce DNS systems. We verify that it is hard for the double-core channel to occur except for extremely high $\beta$. Therefore, we may be able to constrain the value of $\beta$ from observations of intermediate products of each channel.


\section*{Acknowledgement} \label{sec:acknowledgement}
We appreciate the anonymous referee for constructive advice. This work was supported by JSPS KAKENHI grant No. 18J00558 (T.K.), and 17H01130, 17K05380, and 26104007 (H.U.).

\appendix
\section{Full set of the results\label{append1}}
\begin{table*}[b]
 	\caption{The obtained formation rates and merger rates of DNS systems and the rate of USSNe for All 32 models.\label{table:results_full}}
 	\begin{center}
      \begin{tabular}{ccccccccccc}\hline\hline
         Rank & Model & $\beta$ & ``pessimistic"& $\lambda$ & $R_{f,\mathrm{DNS}}$ & $R_{m,\mathrm{DNS}}$ & USSNe & $\log\tilde{\Lambda}$\\
         &&for RLO&CE&&($\pgal~\pmyr$)&($\pgal~\pmyr$)&($\pgal~\pmyr$)&\\\hline
         1 &   7.0.05 & 0.7 & 0 & 0.5 & ~37.67 & ~20.93 & ~555.77~(52.46) & $ 0$\\
         2 &   7.1.05 & 0.7 & 1 & 0.5 & ~33.58 & ~17.27 & ~519.99~(46.99) & $ -0.43$\\
         3 &   7.0.10 & 0.7 & 0 & 1.0 & ~48.17 & ~24.14 & ~572.42~(57.67) & $ -1.62$\\
         4 &      5.0 & 0.5 & 0 & $\lambda_\nj$ & ~12.71 & ~~4.94 & ~510.88~(16.31) & $ -1.75$\\
         5 &      5.1 & 0.5 & 1 & $\lambda_\nj$ & ~12.50 & ~~4.82 & ~488.80~(14.29) & $ -1.89$\\
         6 &   0.0.05 & 0.0 & 0 & 0.5 & 109.72 & ~92.03 & ~744.29~(67.51) & $ -2.74$\\
         7 &      7.1 & 0.7 & 1 & $\lambda_\nj$ & ~17.50 & ~~3.00 & ~755.77~(37.94) & $ -3.27$\\
         8 &      7.0 & 0.7 & 0 & $\lambda_\nj$ & ~18.76 & ~~3.84 & ~785.90~(40.61) & $ -3.31$\\
         9 &   9.0.05 & 0.9 & 0 & 0.5 & ~26.63 & ~10.52 & ~409.08~(25.24) & $ -3.39$\\
         10 &      9.0 & 0.9 & 0 & $\lambda_\nj$ & ~21.51 & ~~7.49 & ~280.67~(12.84) & $ -3.69$\\
         11 &      9.1 & 0.9 & 1 & $\lambda_\nj$ & ~17.54 & ~~5.40 & ~268.87~(11.86) & $ -3.74$\\
         12 &   9.0.10 & 0.9 & 0 & 1.0 & ~28.38 & ~~8.48 & ~258.08~(16.55) & $ -4.26$\\
         13 &      0.0 & 0.0 & 0 & $\lambda_\nj$ & ~~8.51 & ~~4.49 & ~543.11~(20.31) & $ -5.47$\\
         14 &      0.1 & 0.0 & 1 & $\lambda_\nj$ & ~~7.96 & ~~4.04 & ~539.12~(20.13) & $ -5.50$\\
         15 &   0.1.05 & 0.0 & 1 & 0.5 & ~89.35 & ~72.14 & ~677.31~(56.00) & $ -5.62$\\
         16 &   7.1.10 & 0.7 & 1 & 1.0 & ~37.01 & ~13.66 & ~510.09~(44.74) & $ -6.92$\\
         17 &   5.0.05 & 0.5 & 0 & 0.5 & ~62.47 & ~51.50 & ~370.37~(41.44) & $ -7.26$\\
         18 &   5.1.05 & 0.5 & 1 & 0.5 & ~53.64 & ~42.72 & ~305.14~(29.53) & $ -8.29$\\
         19 &   5.0.10 & 0.5 & 0 & 1.0 & ~94.85 & ~66.09 & ~474.28~(55.87) & $ -8.34$\\
         20 &   9.1.05 & 0.9 & 1 & 0.5 & ~18.38 & ~~6.16 & ~249.52~(11.06) & $ -8.77$\\
         21 &   7.1.01 & 0.7 & 1 & 0.1 & ~15.34 & ~~5.71 & ~340.19~(25.70) & $-10.33$\\
         22 &   7.0.01 & 0.7 & 0 & 0.1 & ~18.39 & ~~6.92 & ~364.35~(27.20) & $-10.55$\\
         23 &   5.1.10 & 0.5 & 1 & 1.0 & ~69.79 & ~41.68 & ~379.74~(25.89) & $-10.85$\\
         24 &   9.1.10 & 0.9 & 1 & 1.0 & ~19.42 & ~~3.76 & ~189.24~(~8.98) & $-11.11$\\
         25 &   0.0.01 & 0.0 & 0 & 0.1 & ~20.65 & ~18.77 & ~247.41~(~1.14) & $-13.02$\\
         26 &   0.1.01 & 0.0 & 1 & 0.1 & ~16.06 & ~14.26 & ~234.25~(~0.85) & $-13.19$\\
         27 &   0.0.10 & 0.0 & 0 & 1.0 & 153.73 & 100.96 & 1003.80~(91.25) & $-14.03$\\
         28 &   9.0.01 & 0.9 & 0 & 0.1 & ~13.48 & ~~6.44 & ~274.71~(13.49) & $-15.54$\\
         29 &   0.1.10 & 0.0 & 1 & 1.0 & 107.77 & ~59.18 & ~734.69~(51.39) & $-16.50$\\
         30 &   5.0.01 & 0.5 & 0 & 0.1 & ~11.90 & ~10.85 & ~186.40~(~3.73) & $-22.31$\\
         31 &   5.1.01 & 0.5 & 1 & 0.1 & ~11.88 & ~10.82 & ~186.34~(~3.73) & $-22.31$\\
         32 &   9.1.01 & 0.9 & 1 & 0.1 & ~~9.62 & ~~4.00 & ~263.30~(12.42) & $-22.92$\\\hline
      \end{tabular}
   \end{center}
   \tablecomments{In column 8, the rates given in parentheses are those of iPTF 14gqr like USSNe.}
\end{table*}
The full set of our results is in Table \ref{table:results_full}. We denote the fixed $\lambda$ models as 10X.Y.10Z Here, X=$\beta$, Y=1, or 0 represents if ``pessimistic'' CE is considered or not, and Z=$\lambda$.
\setcounter{table}{5}

\end{document}